\documentclass[11pt]{iopart}

\usepackage{iopams}
\usepackage{cases}
\usepackage{makeidx}
\usepackage{color}
\makeindex
\usepackage {graphicx}
\def\>{\right\rangle}
\def\<{\left\langle}
\def\be{\begin{equation}}
\def\ee{\end{equation}}
\def\ba{\begin{array}{lll}}
\def\ea{\end{array}}

\def\beq{\begin{eqnarray}}
\def\eeq{\end{eqnarray}}
\begin{document}

\title{Non-Abelian BF theory for $2+1$ dimensional topological states of matter}
\author{\bf{A Blasi$^{1,2}$, A Braggio$^{3}$, M Carrega$^{4}$, D Ferraro$^{1,2,3}$, N Maggiore$^{1,2}$ and  N Magnoli$^{1,2}$}}
\address{$^1$ Dipartimento di Fisica, Universit\`a di Genova,Via Dodecaneso 33, 16146, Genova, Italy}
\address{$^2$ INFN, Sezione di Genova, Via Dodecaneso 33, 16146, Genova, Italy}
\address{$^3$ CNR-SPIN, Via Dodecaneso 33, 16146, Genova, Italy}
\address{$^4$ NEST, Istituto Nanoscienze - CNR and Scuola Normale Superiore, I-56126 Pisa, Italy}

\begin{abstract}
We present a field theoretical analysis of the $2+1$ dimensional BF model with boundary 
in the Abelian and the non-Abelian case based on the Symanzik's separability condition. Our aim is to characterize the low energy properties of time reversal invariant topological insulators.
 In both cases, on the edges, we obtain Ka\v{c}--Moody algebras with opposite 
chiralities reflecting the time reversal invariance of the theory. 
While the Abelian case presents an apparent arbitrariness in the value of the central 
charge, the physics on the boundary of the non-Abelian theory is completely determined by time reversal and gauge symmetry. The discussion of the non-Abelian BF model shows that time reversal symmetry on the boundary implies the existence of counter-propagating chiral currents.
 
\end{abstract}
\pacs{11.15.-q, 72.25.-b, 73.43.-f}

\section{Introduction}
During the last decades a new paradigm, based only on the topological properties of a system, has been investigated to classify and predict new states of matter.
The starting point was, in the early 80's, the discovery of the quantum Hall effect (QHE), a phase peculiar of $2+1$ dimensions (D) \cite{Tsui99}, where firstly appeared the connection between a macroscopical physical quantity, i.e. the Hall conductivity, and a class of topological invariants \cite{Thouless82}. This was the first example of new kinds of materials which cannot be described in terms of the principle of broken symmetries.
Recently, new class of systems were discovered which, analogously to the QHE, behave like insulators in the bulk, but support robust conducting edge or surface states, hence displaying non trivial topological properties.
The low energy sector of this kind of states is well described in terms of topological field theories (TFT) \cite{Witten88}. The QHE was successfully described by the Abelian Chern Simons (CS) theory both in the integer and in the fractional regime \cite{Zhang92, Wen95}.
Moreover, the non-Abelian CS theory has been proposed for more exotic fractional states \cite{Wen95}  that are predicted to have excitations with non-Abelian statistics, a key point for the topologically protected quantum computation \cite{Nayak08}. Restriction to the boundary of these TFT has been discussed in order to describe the dynamics of the edge states of the system \cite{Wen95}. 
In contrast with the QHE, where the magnetic field breaks time reversal (T) symmetry, a new class of T invariants systems, i.e. topological insulators (TI), has been predicted \cite{Kane05, Bernevig06a} and experimentally observed \cite{Konig07} in $2+1$ D, leading to the quantum spin Hall effect (QSH). At the boundary of these systems, one has helical states, namely electrons with opposite spin propagating in opposite directions. The presence of these edge currents with opposite chiralities leads to extremely peculiar current-voltage relationship in multi-terminal measurements. The experimental observations carried out by the Molenkamp group \cite{Konig07, Roth09} support these theoretical predictions. Also non trivial realization of TI in $3+1$ D have been conjectured and realized \cite{Qi10, Hazan10}.
States with T and parity (P) symmetry emerge naturally also in lattice models of correlated electrons \cite{Freedman04, Levin05} where doubled CS field theories were developed. In the case of $2+1$ D TI an Abelian doubled CS \cite{Bernevig06b, Levin09} or, equivalently, a BF effective theory has been introduced \cite{Cho11, Santos11}.
The latter TFT allows generalizations to higher dimensions and represents a promising candidate for the description of $3+1$ D TI \cite{Cho11}. Despite the great potential of the BF model in describing T invariants topological states of matter, a careful discussion of these theories in presence of a boundary  is still lacking.

In this paper, we present a detailed and systematic derivation of the helical states at the boundary of the $2+1$ D Abelian  and non-Abelian BF model, applying the Symanzik's method \cite{Symanzik81}. This approach represents the most natural way to introduce a boundary in a quantum field theory and was already considered with success in a different context for the CS case \cite{Emery91, Blasi08, Blasi10}. It is known that $2+1$ D CS theory on a manifold with boundary displays a conformal structure \cite{Moore86}. This is a general property of topological field theories. The BF theory, once a T invariant boundary is introduced, displays two Ka\v{c}--Moody algebras with opposite chiralities, according to the physics of the QSH effect \cite{Bernevig06a, Konig07}, both in the Abelian and in the non-Abelian case  \cite{Maggiore92}.

\section{Abelian BF model}
Here, we recall the main properties of the $2+1$ D Abelian
BF model. Its action depends on two Abelian gauge
fields $A$ and $B$ and, in the Minkowski spacetime, can be written as
\be
S_{bf}= \frac{k}{2\pi}\int d^{3}x \,\epsilon^{\mu \nu\rho}
F_{\mu \nu }B_{\rho} 
\label{BFaction}
\ee
with $F_{\mu \nu}=\partial_{\mu} A_{\nu}- \partial_{\nu} A_{\mu}$ and
$k$ integer due to the invariance of the partition function for large gauge transformations, exactly in the same way as it happens in the CS case \cite{Witten88}. At low energy, the action (\ref{BFaction}) is the only renormalizable one, involving two gauge fields, which is invariant under gauge transformations in $2+1$ dimensions. At this point the T invariance appears to be a discrete symmetry of the theory together with the trivial exchange $A_{\mu}\leftrightarrow B_{\mu}$. In order to describe the model in presence of
a planar boundary it is useful to rewrite (\ref{BFaction}) in the
so called light-cone coordinates $u=y$, $z=(t+x)/\sqrt{2}$ and
$\bar{z}= (t-x)/\sqrt{2}$ where the field components become $(A_{u}, A, \bar{A})$ and $(B_{u}, B, \bar{B})$. In
these new variables the action reads
\be
S_{bf}=\frac{k}{\pi} \int d u d^{2} z\left[B\left( \bar{\partial}
A_{u}-\partial_{u} \bar{A}\right)+\bar{B} \left(\partial_{u} A- \partial A_{u}\right) + B_{u}
\left(\partial \bar{A}- \bar{\partial} A\right)\right].
\ee

The mentioned T symmetry (modulo an exchange of the two gauge fields) in these new variables is
\be
\left(u, z, A, A_{u}, B, B_{u}\right)\leftrightarrow  \left(u,
-\bar{z}, -\bar{A}, A_{u}, \bar{B}, -B_{u}\right);
\ee

notice that the $A$ and $B$  fields transform differently under T, which signals the fact that they can be linked to different physical quantities. Indeed, this allows us to interpret $A$ as a charge density and $B$ as a spin density, the fundamental ingredients of the description of TI \cite{Cho11}. 

As it is well known, gauge field theories are affected by the presence of redundant degrees of freedom, and a gauge fixing choice is necessary in order to deal with the physical degrees of freedom only. The gauge fixing procedure introduces in the theory unphysical ``ghost fields''. Now, while in the Abelian case the ghost fields decouple, and can be integrated out from the Green functions generating functionals, this does not happen in the non-Abelian case, where ghost fields are truly quantum, although unphysical, fields  \cite{Weinberg96}. Once gauge fixed, for the theory  partition function and propagators can be defined, and quantum extension can be discussed. 

Covariance being already broken by the presence of a boundary, it is convenient to adopt the axial gauge choice $A_{u}=B_{u}=0$, and add to (\ref{BFaction}) the corresponding gauge fixing term $S_{gf}$. The coupling with external sources is introduced by means of the term $S_{ext}=\int d u d^{2}z \sum_{\psi} j_{\psi} \psi$, $\psi$ being a generic field of the theory. 

\section{Presence of a boundary}
We consider now the planar boundary $u=0$. The introduction of a boundary in field theory has been thoroughly discussed long time ago by Symanzik in \cite{Symanzik81}. The basic idea is quite simple and general, and concerns the propagators of the theory, which only in the gauge fixed theory are well defined: the propagators of the theory between points lying on opposite sides of the boundary, must vanish. This $separability$ condition results in the following general form of the propagator for any field $\psi$
\be
\Delta^{\psi, \psi'}=\Theta(u) \Theta(u') \Delta^{\psi,
\psi'}_{+}+\Theta(-u) \Theta(-u') \Delta^{\psi, \psi'}_{-}
\ee
where $\pm$ indicates the value of the quantities for
$(u\rightarrow 0^{\pm})$ and $\Theta(u)$ is the Heaviside step function. In addition, the usual field theory constraints of locality and power counting are required together with helicity conservation \cite{Emery91}.

The most general boundary action consistent with the previous
conditions is
\be
S_{bd}=-\frac{k}{\pi} \int  d u d^{2} z \delta(u)
\left(\alpha_{1} A \bar{A}+ \alpha_{2} A \bar{B} + \alpha_{3} \bar{A}
B+ \alpha_{4} B \bar{B}\right),
\label{Action_bd}
\ee
where $\alpha_{1}$, $\alpha_{2}$, $\alpha_{3}$ and $\alpha_{4}$ are free real parameters. The above boundary term leads to $\delta(u)$-dependent breakings of the equations of motion
\beq
\mathcal{F}_{\bar{B}}&=&\sum_{\eta=\pm}\frac{k}{\pi}\delta(u)\left(\alpha_{2}A_{\eta}+\alpha_{4}B_{\eta}\right)
\label{Eq_breaking_in}\\ 
\mathcal{F}_{B}&=&\sum_{\eta=\pm}\frac{k}{\pi}\delta(u)\left(\alpha_{3}\bar{A}_{\eta}+\alpha_{4}\bar{B}_{\eta}\right)\\
\mathcal{F}_{\bar{A}}&=&\sum_{\eta=\pm}\frac{k}{\pi}\delta(u)\left(\alpha_{1}A_{\eta}+\alpha_{3}B_{\eta}\right)\\
\mathcal{F}_{A}&=&\sum_{\eta=\pm}\frac{k}{\pi}\delta(u)\left(\alpha_{1}\bar{A}_{\eta}+\alpha_{2}\bar{B}_{\eta}\right).
\label{Eq_breaking_fin}
\eeq
where $\mathcal{F}_{\psi}=\delta S/ \delta \psi$ and $S= S_{bf}+S_{gf}+S_{ext}$.

From now on, we will focus on the $+$ side of the boundary, the $-$ side being obtained by the P symmetry
$\left(u, z, A, A_{u}, B, B_{u}\right)\leftrightarrow  \left(-u,
\bar{z}, \bar{A}, -A_{u}, \bar{B}, -B_{u}\right)$.

In order to fix the parameters, we integrate (\ref{Eq_breaking_in}-\ref{Eq_breaking_fin}) at vanishing sources with respect to $u$ in the infinitesimal interval
$(-\epsilon, \epsilon)$. This leads to
\beq
(1-\alpha_{2})A_{+}-\alpha_{4}B_{+}&=&0
\label{consistency_in}\\ 
\alpha_{1}A_{+}-(1-\alpha_{3})B_{+}&=&0
\label{consistency_2} \\ 
(1+\alpha_{3})\bar{A}_{+}+\alpha_{4} \bar{B}_{+}&=&0
\label{consistency_3}\\ 
\alpha_{1}\bar{A}_{+}+(1+\alpha_{2}) \bar{B}_{+}&=&0.
\label{consistency_fin}
\eeq

It is easy to verify that (\ref{consistency_in}-\ref{consistency_fin}) have non trivial T invariant
solutions when the determinants vanish, namely
\be
\alpha_{1} \alpha_{4}-(1-\alpha^{2}_{2})=0; \qquad \alpha_{2}=-\alpha_{3}.
\label{contraint}
\ee

The Abelian BF theory with boundary satisfies the following Ward identities (WI) on the generating functional $Z_{c}$:

\beq
&&\frac{\pi}{k}\int du\,H[Z_{c}]\equiv -\frac{\pi}{k}\int du \left(\bar{\partial} j_{\bar{A}}+\partial
j_{A}\right)=\nonumber \\
&&-\left[\alpha_{1} \left(\bar{\partial}A_{+}+\partial
\bar{A}_{+}\right)+\alpha_{2}\left(\partial
\bar{B}_{+}-\bar{\partial} B_{+}\right)\right]
\label{res1}\\
&&\frac{\pi}{k}\int du\,N[Z_{c}]\equiv-\frac{\pi}{k}\int du \left(\bar{\partial} j_{\bar{B}}+\partial
j_{B}\right)= \nonumber\\
&& -\left[\alpha_{2}\left( \bar{\partial}A_{+}-\partial
\bar{A}_{+}\right)+\alpha_{4} \left(\bar{\partial} B_{+}+\partial
\bar{B}_{+}\right)\right].
\label{res2}
\eeq

It is well known that the axial gauge is not a complete gauge fixing \cite{Bassetto91}:  a residual 
gauge invariance remains, expressed by the two WI (\ref{res1}-\ref{res2}), one for each gauge field $A$ and $B$. As it is apparent, the 
presence of the boundary results in $\delta(u)$-dependent $linear$ 
breaking terms on the r.h.s. of (\ref{res1}-\ref{res2}). Such linear terms are 
allowed, since in quantum field theory a 
non-renormalization theorem assures that linear breakings are present 
at classical level only, and do not acquire quantum corrections \cite{Becchi75}. 

In other words, the WI (\ref{res1}-\ref{res2}) represent the most general expression 
of the gauge invariance of the complete theory (bulk and boundary).

\section{Conserved currents and algebra}
We introduce the fields
\beq
R_{+}&\equiv&\left(1-\alpha_{2}\right)
A_{+}+\alpha_{4}B_{+}\\
S_{+}&\equiv&(\alpha_{2}-1)A_{+}+\alpha_{4}B_{+},
\eeq
which, according to (\ref{consistency_in}-\ref{consistency_fin}), satisfy the boundary conditions 
\be
\bar{R}_{+}=S_{+}=0.
\label{annull}
\ee
It is easy to verify that, in terms of these new fields, the WI (\ref{res1}-\ref{res2}) decouple

\beq
\frac{2 \pi}{k} \int d u \left( \bar{\partial} j_{\bar{R}}+ \partial j_{R}\right)&=& \frac{1}{\alpha_{4} (1-\alpha_{2})} \bar{\partial} R_{+}
\label{resR}\\
\frac{2 \pi}{k} \int d u \left( \bar{\partial} j_{\bar{S}}+ \partial j_{S}\right)&=& \frac{1}{\alpha_{4} (1-\alpha_{2})} \partial \bar{S}_{+},
\label{resS}
\eeq
from which we read the chirality conditions
\be
\bar{\partial} R_{+}= \partial \bar{S}_{+}=0.
\label{chiral}
\ee
Notice that $R_{+}$ and $\bar{S}_{+}$ are related by T symmetry and, in virtue of (\ref{annull}) and (\ref{chiral}), are conserved currents. From the WI (\ref{resR}-\ref{resS}), we immediately get the following (Abelian limit of a) Ka\v{c}-Moody algebra

\beq
\left[R_{+}(z), R_{+}(z')\right]&=&\frac{2 \pi \alpha_{4} (1-\alpha_{2})}{k} \partial \delta(z-z')\nonumber \\
\left[ \bar{S}_{+}(\bar{z}), \bar{S}_{+}(\bar{z}')
\right]&=&\frac{2 \pi \alpha_{4} (1-\alpha_{2})}{k}  \bar{\partial}  \delta(\bar{z}- \bar{z}')\nonumber\\
\left[ R_{+}(z), \bar{S}_{+}(\bar{z}')\right]&=&0,
\label{Abelian_algebra}
\eeq
that are connected by T and where the last commutator shows the decoupling of the two currents.
We stress that, in the Abelian case, the central charge is not completely fixed since it depends on the boundary parameters. Once required the decoupling of the boundary action (\ref{Action_bd}) in terms of $R_{+}$ and $S_{+}$ we can fix the values of the parameters $\alpha_{2}=0$ and $\alpha_{4}=1$.

The result of our analysis is the presence, on the boundary of the
Abelian BF model, of two conserved currents with opposite chiralities,
connected by T symmetry. The above picture is
in accordance with the phenomenology involved in the helical edge
states of the QSH effect \cite{Bernevig06a, Konig07, Qi10} in agreement with the idea that the BF model  is
a good effective field theoretical description for the $2+1$ D TI \cite{Cho11}. Note that, due to
the presence of the factor $k$, our results could be applied to possible
fractional extension as well \cite{Bernevig06b}.

\section{Non-Abelian BF model}
The non-Abelian generalization of (\ref{BFaction}) is
\begin{equation}
S_{BF}=\frac{k}{2\pi}\int\!\! d^3x\ \varepsilon^{\mu\nu\rho}\left\{
F^a_{\mu\nu}B^a_\rho+\frac{1}{3}f^{abc}B^a_\mu
B^b_\nu B^c_\rho\right\}
\ ,
\label{nonabbf}\end{equation}
where the fields $A_{\mu}^a$ and $B_{\mu}^a$ belong to the adjoint 
representation of a compact simple gauge group $G$ whose structure 
constants are $f^{abc}$. In (\ref{nonabbf}), the non-Abelian
field 
strength is defined as 
$F^a_{\mu\nu} = \partial_\mu A^a_\nu - \partial_\nu A^a_\mu
+ f^{abc} A^b_\mu A^c_\nu$ and the coupling $k$ can be related to the mutual statistics between the quasiparticle sources of the fields. It is worth to note that $k$ can be connected to the "cosmological constant" usually defined in literature \cite{Maggiore92}.

In light--cone coordinates the action in (\ref{nonabbf}) reads
\begin{eqnarray}
S_{BF}&=&\frac{k}{\pi}\int du d^2z \{ [
B^a(\bar{\partial}A_u^a-\partial_u\bar{A}^a+
f^{abc}\bar{A}^b A_u^c)+\bar{B}^a (\partial_uA^a-\partial A_u^a+f^{abc}
A_u^bA^c) \nonumber\\
&& +B_u^a(\partial\bar{A}^a-\bar{\partial}A^a
+f^{abc}A^b\bar{A}^c)]+ f^{abc}B^a\bar{B}^bB_u^c \}\ .
\label{lcaction}
\end{eqnarray}
The axial gauge $A^a_u=B^a_u=0$ is realized, as in the Abelian case, through a suitable 
gauge fixing term which involves ghost, antighost and lagrange 
multipliers for both the gauge fields $A_{\mu}^a$ and $B_{\mu}^a$.  
It is straightforward to verify that the above action satisfies T and P symmetries. The non-Abelian gauge symmetry is realized, as usual in the 
axial gauge,  by means of the local WI $H^a[Z_{c}] =  N^a[Z_{c}] = 0$, 
where $H^a(u,z,\bar{z})$ and $N^a(u,z,\bar{z})$ are local operators
whose explicit form 
is unessential here, and can be found in \cite{Maggiore92},
where the non-Abelian BF 
theory in the axial gauge with and without boundary is treated in 
greater detail.

As in the Abelian case, the introduction of the planar boundary $u=0$
is realized through the 
addition of a term in the action or, equivalently, through
$\delta(u)$ - breakings of the fields equations, satisfying the 
constraints of consistency, power counting and helicity conservation. 
Consequently, the local WI are broken by the boundary and, once integrated over the $u$-coordinate, read
\begin{eqnarray}
\frac{\pi}{k}\int  du\,H^a [Z_c]
&=&\Delta^{a}_{H}(z,\bar{z})
\label{brokenward1}\\
\frac{\pi}{k} \int du\,N^a [Z_c]
&=&\Delta^{a}_{N}(z,\bar{z}).
\label{brokenward2}
\end{eqnarray}
The above equations are the non-Abelian counterpart of (\ref{res1}-\ref{res2}), and likewise they describe the 
residual gauge invariance left on the boundary. By choosing the parameters $\alpha_{1}=\alpha_{4}$ and $\alpha_{2}=\alpha_{3}$ the terms 
$\Delta^{a}_{H}$ and $\Delta^{a}_{N}$ on their r.h.s. are linear  in the quantum fields
(hence classical and allowed). 
Notice that the WI (\ref{brokenward1}-\ref{brokenward2}) are the most general form of gauge invariance for the non-Abelian theory with boundary \cite{Becchi75}. This is a central point for what 
follows. Requiring T symmetry at the boundary and (\ref{contraint}), still valid in the non-Abelian case, one fixes the parameters of the boundary action to be $\alpha_{1}=\alpha_{4}=1$ and $\alpha_{2}=\alpha_{3}=0$.

The most general conserved chiral currents
on the 
boundary, together with their algebra, can be fully determined \cite{Maggiore92}. It turns out
that the only T symmetric solution is given by a  direct sum of two independent Ka\v{c}-Moody algebras satisfied by conserved currents of {\it{opposite chirality}}\footnote{Note that other solutions exist that break T symmetry at the boundary, but we will not consider them here.}.

In more detail, the WI corresponding to this solution, read:

\begin{eqnarray}
\frac{\pi}{k}\int du\,H^a [Z_c] &=&
-(\partial\bar{A}^a_{+}+\bar{\partial}A^a_{+})\label{lw1}\\
\frac{\pi}{k}\int du\,N^a [Z_c] &=&
-(\partial\bar{B}^a_{+}+\bar{\partial}B^a_{+}) .
\label{lw2}
\end{eqnarray}

The linear combinations of fields 
\be 
R^a_{+} \equiv  A^a_{+}+ B^a_{+}\quad\quad 
S^a_{+} \equiv - A^a_{+} +B^a_{+}
\label{redef}\ee
satisfy 
Dirichlet boundary conditions 
$S^a_+(z,\bar{z})=\bar{R}^a_+(z,\bar{z})=0$, and the residual WI (\ref{lw1}-\ref{lw2}) identify two conserved currents with opposite chiralities
$\bar{\partial}R^a_{+}(z,\bar{z})=\partial\bar{S}^a_{+}
(z,\bar{z})=0$, for which the following direct sum of Ka\v c-Moody algebras 
living on the same side of the plane $u=0$ with 
opposite chiralities, holds:
\begin{eqnarray}
&&\left [R^a_{+}(z),R^b_{+}(z')\right]=
f^{abc}\delta(z-z')R^c_{+}(z)+\frac{2\pi}{k}\delta^{ab}\delta'(z-z')\nonumber\\
&&\left [\bar{S}^a_{+} ( \bar{z} ) , \bar{S}^b_{+} ( \bar{z} ' ) \right 
] =
f^{abc}\delta(\bar{z}-\bar{z}')\bar{S}^c_{+}(\bar{z})+\frac{2\pi}{k}\delta^{ab}
\delta'(\bar{z}-\bar{z} ')\nonumber\\
&&\left [ R^a_{+}({z}),\bar{S}^b_{+}(\bar{z} ')\right ]=0
\end{eqnarray}
where the last commutator shows that $R_{+}$ and $\bar{S}_{+}$ are decoupled.
Note that, in the Abelian limit, the above WI  and algebras reduce to (\ref{res1}-\ref{res2}) and  (\ref{Abelian_algebra}) respectively in the case of decoupling of the boundary action.

Finally, we notice that, as it has been shown long time ago 
in \cite{Guadagnini:1990aw}, the redefinition (\ref{redef}) allows to rephrase the $2+1$ D non-Abelian BF theory in (\ref{nonabbf}) in terms 
of two CS theories with opposite coupling constants in the bulk. As stated before, in the non-Abelian case, the consistency of the theory  also requires the complete decoupling at the boundary that, conversely, is not needed in the Abelian case.

\section{Conclusions}
We investigated the $2+1$ D BF model in presence of a boundary both in the Abelian and in the non-Abelian case as a proper effective field theory for the QSH states. The key 
points of our analysis have been the combined application of the Symanzik's separability condition and the T symmetry. By means of these conditions we have been able to evaluate the algebraic structure of the current at the edge of the system. In the Abelian case we found two Ka\v{c}--Moody current algebras with opposite chiralities and with the central charges depending on the coupling constant $k$, together with arbitrary boundary parameters. In the non-Abelian case we also find two  Ka\v{c}--Moody currents with opposite chiralities with the important difference that the central charge is unambigously determined in terms of the coupling constant $k$ of the theory. The appearance of Ka\v{c}--Moody algebras related by T symmetry, reflects the equivalence between the double CS model and the non-Abelian BF model once the two CS are completely decoupled at the boundary \cite{Guadagnini:1990aw}.  The QSH is characterized by the presence of currents with opposite spin and chiralities on the boundary \cite{Konig07, Roth09, Qi10}. One of the new results of this paper is that we find an algebraic structure which displays this features and, in addition, it respects the T invariance. 

The Symanzik's method of treating boundaries in field theory is widely general and standard, and it is model independent. It is precisely for this reason that we adopted it, having in mind two recently and widely discussed generalizations: the $3+1$ D case and the breaking of the T symmetry, which can be easily treated along the same lines illustrated in this paper.

\section*{Acknowledgements}
We thank M. Sassetti for useful discussions. We acknowledge the support of the EU-FP7 via ITN-2008-234970 NANOCTM.


\section*{References}
\begin{thebibliography}{10}
\bibitem{Tsui99} Tsui D C 1999 Rev. Mod. Phys. \textbf{71} 891
\bibitem{Thouless82} Thouless D J, Kohmoto M, Nightingale M P and den Nijs M 1982 Phys. Rev. Lett. \textbf{49} 405
\bibitem{Witten88} Witten E 1988 Commun. Math. Phys. \textbf{117} 353
\bibitem{Zhang92} Zhang S C 1992 Int. J. Mod. Phys. B \textbf{6} 25
\bibitem{Wen95} Wen X G 1995 Adv. Phys. \textbf{44} 405
\bibitem{Nayak08} Nayak C, Simon S H, Stern A, Freedman M and  Das Sarma S 2008 Rev. Mod. Phys. \textbf{80} 1083
\bibitem{Kane05} Kane C L and  Mele E J 2005 Phys. Rev. Lett. \textbf{95} 146802
\bibitem{Bernevig06a} Bernevig B A, Hughes T L and Zhang S C 2006 Science \textbf{314} 1757
\bibitem{Konig07} Konig M, Weidmann S, Brune C, Roth A, Buhmann H, Molenkampf L W, Qi X L and Zhang S C 2007 Science \textbf{318} 766
\bibitem{Roth09} Roth A, BrŸne C, Buhmann H, Molenkamp L W, Maciejko J, Qi X L and Zhang S C 2009 Science \textbf{325} 294
\bibitem{Qi10} Qi X L and Zhang S C 2011 Rev. Mod. Phys. \textbf{83} 1057
\bibitem{Hazan10} Hasan M Z and Kane C L 2010 Rev. Mod. Phys. \textbf{82} 3045
\bibitem{Freedman04} Freedman M, Nayak C, Shtengel K, Walker K and Wang Z 2004 Ann. Phys. {\bf 310} 428
\bibitem{Levin05} Levin M A and Wen X G 2005 Phys. Rev. B \textbf{71} 045110
\bibitem{Bernevig06b} Bernevig B A and Zhang S C 2006 Phys. Rev. Lett. \textbf{96} 106802
\bibitem{Levin09} Levin M A and Stern A 2009 Phys. Rev. Lett. \textbf{103} 196803
\bibitem{Cho11} Cho G Y and Moore J E 2011 Ann. Phys. \textbf{326} 1515
\bibitem{Santos11} Santos L, Neupert T, Ryu S, Chamon C and Mudry C 2011 arXiv:1108.2440v1
\bibitem{Symanzik81} Symanzik K 1981 Nucl. Phys. B \textbf{190} 1
\bibitem{Emery91}  Emery S and Piguet O 1991 Helv. Phys. Acta \textbf{64}  1256
\bibitem{Blasi08} Blasi A, Ferraro D, Maggiore N, Magnoli N, and Sassetti M 2008 Ann. der Phys. \textbf{17} 885 
\bibitem{Blasi10} Blasi A, Maggiore N, Magnoli N and Storace S 2010 Class. Quantum. Grav. \textbf{27} 165018
\bibitem{Moore86} Moore G W and Seiberg N 1989 Phys. Lett. B \textbf{220} 422
\bibitem{Maggiore92} Maggiore N and Provero P 1992 Helv. Phys. Acta \textbf{65} 993
\bibitem{Weinberg96} Weinberg S 1996 \emph{The quantum theory of fields}  (United Kingdom: Cambridge University Press)
\bibitem{Bassetto91} Bassetto A, Nardelli G and Soldati R 1991 \emph{Yang-Mills theories in algebraic noncovariant gauges} (Singapore: World Scientific) 
\bibitem{Becchi75} Becchi C, Rouet A and Stora R 1975 (Erice: Ettore Majorana Summer School) 
\bibitem{Guadagnini:1990aw} Guadagnini E, Maggiore N and Sorella S P 1990 Phys. Lett.  B \textbf{247} 543
\end{thebibliography}
\end{document}